\documentclass[11pt]{article}
\usepackage{amsmath,amssymb,color,graphics,epsfig,cite}

\usepackage{amsfonts}

\makeatletter
\@addtoreset{equation}{section}
\makeatother

\newcommand{\be}{\begin{equation}}
\newcommand{\ee}{\end{equation}}
\newcommand{\bea}{\setlength\arraycolsep{2pt} \begin{eqnarray}}
\newcommand{\eea}{\end{eqnarray}}
\newcommand{\nn}{\nonumber}

\def\ft#1#2{{\textstyle{\frac{\scriptstyle #1}{\scriptstyle #2} } }}
\def\fft#1#2{{\frac{#1}{#2}}}

\def\0{{\sst{(0)}}}
\def\1{{\sst{(1)}}}
\def\2{{\sst{(2)}}}
\def\3{{\sst{(3)}}}
\def\4{{\sst{(4)}}}
\def\5{{\sst{(5)}}}
\def\6{{\sst{(6)}}}
\def\7{{\sst{(7)}}}
\def\8{{\sst{(8)}}}
\def\sst#1{{\scriptscriptstyle #1}}

\begin{document}

\begin{center}
{\Large {\bf Vacua and Exact Solutions in Lower-$D$ Limits of EGB}}

\vspace{20pt}

{\large Liang Ma and H. L\"u}

\vspace{10pt}

{\it  Center for Joint Quantum Studies and Department of Physics,\\
School of Science, Tianjin University, Tianjin 300350, China}

\vspace{40pt}

\underline{ABSTRACT}
\end{center}

We consider the action principles that are the lower dimensional limits of the Einstein-Gauss-Bonnet gravity {\it via} the Kaluza-Klein route. We study the vacua and obtain some exact solutions.  We find that the reality condition of the theories may select one vacuum over the other from the two vacua that typically arise in Einstein-Gauss-Bonnet gravity. We obtain exact black hole and cosmological solutions carrying scalar hair, including scalar hairy BTZ black holes with both mass and angular momentum turned on.  We also discuss the holographic central charges in the asymptotic AdS backgrounds.



\thispagestyle{empty}
\pagebreak

\section{Introduction}

Einstein theory of gravity admits naturally higher-derivative extensions by appending the Riemann tensor polynomial invariants to the Einstein-Hilbert action. The field equations in general involve four derivatives, leading to possible renormalizablity; however, the theory contains an inevitable ghostlike massive graviton \cite{Stelle:1976gc}.  There exist specific combinations of tensor polynomials, such as the Gauss-Bonnet term or the more general Lovelock series, that give rise to field equations of second order derivatives \cite{L1}. Unfortunately, these terms either vanish or become total derivative in four or lower dimensions. In four dimensions, when an extra scalar is available, analogous constructions can also be performed such that the field equations remain two derivatives. The resulting theories are Horndeski \cite{H} or the more general Galilean \cite{Deffayet:2009mn} gravities. In fact the Kaluza-Klein reduction of Einstein-Gauss-Bonnet (EGB) gravity leads naturally to a class of Hordenski theories \cite{VanAcoleyen:2011mj}.

Recently the $D\rightarrow 4$ limit of Einstein-Gauss-Bonnet (EGB) gravity was proposed by first rescaling the coupling constant $\alpha$ with $\alpha/(D-4)$ \cite{GL}.  However, the claim that the resulting theory is of pure graviton was cast in doubt by several grounds. Intuitively this is not possible since the Gauss-Bonnet term is the only quadratic curvature invariant with two-derivative field equations. Indeed the Kaluza-Klein approach of the $D\rightarrow 4$ limit leads to a class of scalar-tensor theory that belongs to the Horndeski class \cite{lp}. (See also \cite{Kobayashi:2020wqy}.) Inconsistency in the covariant approach was shown in \cite{Gurses:2020ofy}. (See also \cite{Ai:2020peo,Shu:2020cjw}.) These demonstrate that the $D\rightarrow 4$ limit of EGB gravity could only be done in some specific backgrounds. Indeed, it turns out that this limit yields some finite result in a variety of backgrounds, see {\it e.g.}~\cite{backgrounds}. However, not all the backgrounds are equivalent and we would naturally want the four-dimensional covariance without insisting on the full covariance in general dimensions, if our goal is to obtain a covariant four-dimensional theory that is the $D\rightarrow 4$ limit of EGB.  This belongs precisely to the Kaluza-Klein idea and it is consistent to keep only the breathing mode of the internal space together with the four-dimensional metric.  There exists then a smooth $D\rightarrow 4$ limit of the resulting Kaluza-Klein theory \cite{lp} while keeping some nontrivial aspects of the higher-dimensional EGB.  The resulting theory belongs to a special class of Horndeski gravity, consists of the metric and a scalar as its minimum field content.  Analogous approach was also employed for Einstein gravity in the $D\rightarrow 2$ limit \cite{2d1}. (See also recent works \cite{2d2,Ai:2020peo,eoms,Hennigar:2020lsl}.) 

It should be emphasized that after taking the $D\rightarrow 4$ limit, the theory is then intrinsically four dimensional. This is different from the previous Kaluza-Klein approach which are simply the zero modes of an intrinsic higher-dimensional theory.  In fact before taking the limit, the Kaluza-Klein theories in different conformal frames associated with the scalar field are (at least locally) equivalent, but the singular $D\rightarrow 4$ limit will freeze the conformal frames and lead to inequivalent theories, including those constructed in \cite{lp} and also one that contains a minimally coupled scalar, but with the Gauss-Bonnet term dropped \cite{Bonifacio:2020vbk}.

The Kaluza-Klein approach selects a specific class of backgrounds that retain the lower-dimensional diffeormorphism, but not the full general covariance of the higher-dimensional theory. This implies that some of the $D\rightarrow 4$ limit of EGB gravity may not survive in the Kaluza-Klein approach.  In particular, the limit {\it via} the Kaluza-Klein route may select only one of the two vacua of the higher dimensional analogue.  It is well understood that EGB gravity admits two maximally symmetric vacua of cosmological constant $\lambda_D$, satisfying the quadratic constraint
\be
\Lambda_0=\ft12 (D-1)(D-2) \Big(\lambda_D +\alpha (D-3)(D-4) \lambda_D^2\Big)\,,
\ee
where $\Lambda_0$ is the bare cosmological constant and $\alpha$ is the coupling of the Gauss-Bonnet term. Following the redefinition $\alpha\rightarrow \alpha/(D-4)$ and then taking $D\rightarrow 4$, one arrives at
\be
\Lambda_0 = 3 \lambda_4(1 + \alpha \lambda_4)\,,\label{p4vacuum0}
\ee
which implies that there are two maximally symmetric vacua in the four-dimensional theory, with the effective cosmological constants
\be
\lambda_4^\pm =\fft{1}{2\alpha} \Big( - 1 \pm \sqrt{1 + \ft43 \alpha \Lambda_0}\Big)\,.\label{p4vacuapm}
\ee
However, the scalar-tensor theory obtained in \cite{lp} will select only one preferred vacuum when the effective cosmological constants $\lambda_4^\pm$ of the two vacua have the opposite signs. This may avoid the subtle issue of vacuum collapsing observed in \cite{Shu:2020cjw} for the general $D\rightarrow 4$ limit.

It turns out that the Kaluza-Klein approach to EGB can not only yield a four-dimensional theory, but also three classes of further lower-dimensional theories. The purpose of this paper is to study the vacua of these theories and also construct some exact solutions. The paper is organized as follows.  In sections 2,3,4, we study the vacua and construct exact solutions for the $D\rightarrow p$, $D\rightarrow p+1$ and $D\rightarrow p+2$ theories constructed in \cite{lp}.  In section 5, we study some holographic properties of the $D\rightarrow p$ theory for $p=4$.  We conclude the paper in section 6.

\section{The $D\rightarrow p$ theory}

The theory was obtained by first compactifying EGB gravity on a $(D-p)$-dimensional maximally symmetric space of curvature $\lambda$ and then taking the $D\rightarrow p$ limit, with $p\le 4$. The $D\rightarrow p+1$ and $D\rightarrow p+2$ limits will be discussed in the subsequent sections. The Lagrangian involving the metric and the breathing mode is \cite{lp}
\bea
{\cal L}_p &=& \sqrt{-g} \Big[R
-2\Lambda_0+\alpha\Big(\phi\, E^{\rm GB}+4G^{\mu\nu}\partial_\mu\phi\partial_\nu\phi
-2\lambda Re^{-2\phi}-4(\partial\phi)^2\Box\phi\nn\\
&&+2((\partial\phi)^2)^2-12\lambda(\partial\phi)^2e^{-2\phi}-6\lambda^2e^{-4\phi}\Big)\Big],\label{theory1}
\eea
where $E^{\rm GB} = R^2 - 4 R^{\mu\nu} R_{\mu\nu} + R^{\mu\nu\rho\sigma} R_{\mu\nu\rho\sigma}$. We find that the Gibbons-Hawking surface term is
\bea
S_{\rm surf}&=&2\int d^{p-1} x \sqrt{-h} \Big( (1-2\alpha\lambda e^{-2\phi}) K - \ft23\alpha \phi(K^3 - 3 K K^{(2)} + 2K^{(3)})\nn\\
&&+2\alpha \big(K^{\mu\nu} \partial_\mu \phi\partial_\nu \phi + K n^\mu n^\nu
\partial_\mu \phi \partial_\nu \phi - K (\partial\phi)^2\big)\Big),\label{surf1}
\eea
where $n^\mu$ is the unit vector normal to the surface and $K$ is the trace of the second fundamental form
$K_{\mu\nu}=h_{\mu}{}^\rho \nabla_\rho n_\nu$ and $h_{\mu\nu} = g_{\mu\nu} - n_\mu n_\nu$, and $K^{(2)} = K^{\mu}{}_\nu K^{\nu}{}_{\mu}$ and $K^{(3)} = K^{\mu}{}_\nu K^{\nu}{}_{\rho} K^{\rho}{}_{\mu}$.  The first term in (\ref{surf1}) follows from that in Einstein-Hilbert action \cite{Gibbons:1976ue}.  The second term is associated with the Gauss-Bonnet combination \cite{Liu:2008zf}.  The third one is the surface term for the Horndeski structure $G^{\mu\nu} \partial_\mu \phi \partial_\nu \phi$ \cite{Li:2018rgn}.

The theory contains three nontrivial couplings, $\Lambda_0$, $\alpha$ and $\lambda$, and it is independent of the dimension $p$.  The covariant equations of motion are presented in appendix \ref{app:dptheory}. The maximally symmetric vacuum with the Riemann tensor and constant scalar in $p$ dimensions
\be
R_{\mu\nu\rho\sigma} = \lambda_p\,(g_{\mu\rho} g_{\nu\sigma} - g_{\mu\sigma} g_{\nu\rho})\,,\qquad
\phi=\phi_0\,,
\ee
is specified by
\bea
&&\alpha \Big(p(p-1)(p-2)(p-3)\lambda_p^2 + 4 p(p-1) \lambda_p\lambda\,e^{-2\phi_0}
+24\lambda^2 e^{-4\phi_0}\Big)=0\,,\nn\\
&&\Lambda_0 = \ft12 (p-1)(p-2) \lambda_p +\alpha \Big(\ft12(p-1)(p-2)(p-3)(p-4) \phi_0 \lambda_p^2\nn\\
&&\qquad\qquad - (p-1)(p-2) \lambda\lambda_p e^{-2\phi_0} - 3\lambda^2 e^{-4\phi_0}\Big)\,.
\eea
Unless for $\lambda_p=0=\lambda$, the reality condition for the first equation above requires that $p=4,3,2$.

\subsection{$p=4$}

A particularly noteworthy property in $p=4$ dimensions is that scalar equation and the trace of Einstein equation lead to a geometric constraint on the curvature
\be
R+ \ft12 \alpha L_{\rm GB}=4\Lambda_0\,.\label{p4cons}
\ee
This is the direct $D\rightarrow 4$ limit of the trace equation of EGB gravity, and may be viewed as a litmus test of the $D\rightarrow 4$ limit of EGB. It should be understood that the equation (\ref{p4cons}) is necessary, but not sufficient. Not all the solutions of (\ref{p4cons}) can solve the full set of equations of motion. Nevertheless, for metric Ans\"atze involving only one unknown function, this equation becomes particularly useful.

\subsubsection{Vacuum and the linear spectrum}

The maximally symmetric vacuum is given by
\be
\Lambda_0=3\lambda_4(1+ \alpha \lambda_4)\,,\qquad \lambda_4=-\lambda e^{-2\phi_0}\,.\label{p4scalarvac}
\ee
Note that $(\Lambda_0, \lambda, \alpha)$ are coupling constants of the theory whilst $(\lambda_4,\phi_0)$ are the integration constant characterizing the vacuum. The first equation directly reproduce the $D\rightarrow 4$ limit discussed in the introduction; however, we have now an extra constraint from the scalar. The reality condition requires that $\lambda_4$ has the opposite sign of $\lambda$.  This implies that when the two roots of the first equation have opposite signs, only one vacuum will be selected.  To be specific, we consider $\alpha>0$ and it follows from (\ref{p4vacuapm}) that we must have $\alpha \Lambda_0\ge -3/4$ for the existence of maximally symmetric vacua. When the inequality is saturated, $\lambda_4^\pm$ coalesce, giving rise to a gravity theory without graviton \cite{Fan:2016zfs}.  In the region $-3/4 <\alpha\Lambda_0<0$, both $\lambda^\pm_4$ are negative and both solutions are selected or rejected by the reality condition of the scalar equation (\ref{p4scalarvac}).  When $\alpha\Lambda_0>0$, we have $\lambda_4^-<0$, and it becomes the vacuum of the theory with positive $\lambda$. On the other hand, $\lambda_4^+>0$, and it becomes the vacuum of the theory with negative $\lambda$. In other words, each theory with specific $\lambda$ can only admits one specific vacuum, rather than two. The Minkowski vacuum arises when both $\Lambda_0$ and $\lambda$ vanish, in which case, the Minkowski is the only vacuum of the theory.

Having obtained the vacuum of the theory, we can analyse the linear spectrum by considering the general linear perturbations
\be
g_{\mu\nu} = \bar g_{\mu\nu} + h_{\mu\nu}(x)\,,\qquad \phi = \phi_0 + \varphi(x)\,.
\ee
We find that the linear spectrum contains only the graviton, described by the transverse and traceless $h_{ij}$, satisfying
\be
\kappa_{\rm eff} (\Box - 2\lambda_4) h_{ij}=0\,,\qquad \kappa_{\rm eff} = 1 - 2\alpha \lambda e^{-2\phi_0}=
1+2\alpha \lambda_4\,.\label{p4kappaeff}
\ee
The overall constant $kappa_{\rm eff}$ is the inverse of the effective Newton constant.  For $\lambda^\pm_4$
given in (\ref{p4vacuapm}), we have
\be
\kappa_{\rm eff}^\pm = \pm \sqrt{1 + \ft43\alpha \Lambda_0}\,.
\ee
Thus we see that the ghost free condition selects the $\lambda_4^+$ vacuum.  This implies that for $\alpha\Lambda_0>0$, the $\lambda<0$ theory is unitary, whilst the $\lambda>0$ theory is not.  The Minkowski spacetime as the vacuum of the $\Lambda_0=0=\lambda$ theory is unitary.

Although the theory contains a scalar field, it has no propagating degree of freedom. The would-be scalar equation takes the form
\be
\kappa_\phi (\Box - 4\lambda_4) \varphi =0\,,\qquad\kappa_{\phi} = 12 (\lambda_4 + \lambda e^{-2\phi_0})=0\,.
\ee
In other words, the coefficient $\kappa_\phi$ vanishes identically for {\it all} the maximally symmetric vacua.  Note that had $\kappa_\phi\ne 0$, the scalar perturbation is massive, which is consistent with the fact that the breathing mode of non-flat internal space is typically massive.  Thus we see that although the scalar is indispensable in the four dimensional theory, the linear perturbation of the vacuum nevertheless contains only the graviton.  It is therefore not entirely unreasonable to state that the novel $D=4$ EGB theory is a theory of graviton, as claimed in \cite{GL}.

\subsubsection{Static solutions}

We consider a general class of ansatz of the type
\be
ds^2=-e^{-2\chi(r)} f(r) dt^2 + \fft{dr^2}{f(r)} + r^2 \Big(\fft{dx^2}{1-k x^2} + x^2 d\varphi^2\Big)\,,\qquad
\phi=\phi(r)\,.
\ee
The effective Lagrangian was obtained, given by \cite{lp}
\bea
L_{\rm eff} &=&e^{-\chi}\Big[2(k-\Lambda_0 r^2 - f- r f') +\ft23\alpha\phi'\Big(
3 r^2 f^2 \phi '^3+2 r f \phi '^2 \left(-r f'+2 r f \chi '-4 f\right)\nn\\
&&-6 f \phi ' \left(-r f'+2 r f \chi '-f+k\right)-6 (f-k) \left(f'-2 f \chi '\right)\Big)\nn\\
&& + 4\alpha \lambda e^{-2\phi} \Big(r^2 f' \phi '-2 r^2 f \chi ' \phi '-3 r^2 f \phi '^2+r f'+f-k\Big)-6\alpha\lambda^2 r^2 e^{-4\phi}\Big]\,.
\eea
Here we shall focus on the $\chi=0$ solutions, in which case, the $f$ equation is
\be
\alpha \Big((r\phi'-1)^2 f -\lambda r^2 e^{-2\phi} -k\Big)(\phi'' + \phi'^2)=0\,.
\ee
Thus we see there are two branches of solutions. The branch associated with vanishing
of the first bracket, namely
\be
(r\phi'-1)^2 f= \lambda r^2 e^{-2\phi} +k\,,\label{p4scalareom1}
\ee
was studied in \cite{lp}.  It gives rise to the metric that is the $D\rightarrow 4$ limit of the EGB black holes.
\be
f_\pm =k + \fft{r^2}{2\alpha} \Big(1 \pm \sqrt{1 + \ft43\alpha \Lambda_0 + \fft{8\alpha M}{r^3}}\Big)\,.
\ee
The existence of this black hole can be viewed as the litmus test of the lower-dimensional limit of EGB gravity. An important difference however is that the black hole here carries nontrivial scalar hair, determined by (\ref{p4scalareom1}), and the scalar solution is \cite{lp}
\be
\label{p4scalarsol}
\phi_\pm = \log \ft{r}{L} +\log \Big(\cosh (\sqrt{k}\,\psi) \pm \sqrt{1+\lambda\,L^2k^{-1}} \sinh(\sqrt{k}\,\psi)\Big)\,,\quad \psi = \int_{r_+}^r \fft{du}{u\sqrt{ f(u)}}\,,
\ee
where $L$ is an arbitrary integration constant.  As was emphasized in \cite{lp} that the scalar field is crucial to reproduce the entropy of the black hole that is indeed the $D\rightarrow 4$ limit of the entropy of the black holes in EGB gravity.  This is indicative that the lower dimensional theory cannot be simply a metric theory.

The other branch, corresponding to $\phi'' + \phi'^2=0$, leads to a solution
\be
f=\fft{r^2}{\ell^2}\,,\qquad \phi=\log \Big(\sqrt{\ell^2 \lambda}\,\big(\fft{r}{r_0}-1\big)\Big)\,,\qquad
\Lambda=-3\ell^{-2} + 3\alpha \ell^{-4}\,,\qquad k=0\,.
\ee
The metric describes the AdS spacetime in Poincar\'e coordinates, but the full conformal symmetry is broken to only the Poincar\'e symmetry by the scalar field.  We may refer to such solution as the nearly AdS vacuum.
Such solutions are common in Hordenski gravity and they may be dual to scale invariant quantum field theories that are no conformal \cite{Li:2018rgn}.

\subsubsection{Cosmological solutions}

It is also of interest to study the cosmological solutions of this theory.  The FLRW ansatz is
\be
ds_4=-dt^2 + a(t)^2 (dx_1^2 + dx_2^2 + dx_3^2)\,,\qquad \phi=\phi(t)\,.
\ee
The cosmological equations are
\bea
\dot H  &=&  \fft{2(\Lambda_0 -3 H^2 -3\alpha H^4)}{3(1 + 2\alpha H^2)}\,,\qquad H\equiv \fft{\dot a}{a}\,,\nn\\
\ddot \phi &=& \fft{1}{3(1 + 2\alpha H^2)}\Big(2 \Lambda _0+3 \lambda  e^{-2 \phi }-3 H^2 (1-2 \alpha  \lambda e^{-2 \phi })-3 H (1+2 \alpha  H^2) \dot\phi
\Big)\,,\label{p4cosmoeom}
\eea
together with the Hamiltonian constraint
\be
{\cal H}=2a^3 (\Lambda_0 - 3H^2)+6\alpha a^3(\dot \phi^2 - 2H \dot \phi +\lambda e^{-2\phi})
(\dot \phi^2 - 2H \dot \phi + 2H^2 + \lambda e^{-2\phi})=0\,.
\ee
It was observed in \cite{Kobayashi:2020wqy} that when $\lambda=0$, the scalar equation in (\ref{p4cosmoeom}) can be solved simply by $\dot\phi=H$.  For general $\lambda$, we find a simple solution of de Sitter metric with $H$ being constant, given by
\be
a=e^{H t}\,,\qquad \phi = \log \left(e^{H t} + \sqrt{-\ft{\lambda}{H^2}}\right)\,,
\qquad\Lambda_0=3 H^2 \left(1+\alpha  H^2\right)\,.
\ee
The reality condition implies that $\lambda\le 0$. The solution runs from a de Sitter vacuum at $t\rightarrow -\infty$ to a nearly de Sitter spacetime at large $t$ with linear time-dependence of the scalar $\phi\sim t$.

\subsection{$p=3$}

We now consider the Lagrangian (\ref{theory1}) in $p=3$ dimensions. The maximally symmetric vacuum is give by
\be
\Lambda_0 = \lambda_3 (1 - \alpha \lambda_3)\,, \qquad \lambda e^{-2\phi_0} = - \lambda_3\,.
\ee
Note that if the two roots of 3-dimensional effective cosmological constant $\lambda_3$ from the first equation have opposite signs, only one vacuum will be selected by the constraint of the second equation.
The effective $\kappa$ for the graviton is
\be
\kappa_{\rm eff}=1 + 2\alpha \lambda_3\,.
\ee
There is a kinetic term of the linearized scalar perturbation and the coefficient of the kinetic term is
\be
\kappa_\phi=8\alpha \lambda_3\,.
\ee
Note that the scalar kinetic term vanishes in Minkowski vacuum with $\lambda_3=0$.

For the special static ansatz
\be
ds_3^2 = - f dt^2 + \fft{dr^2}{f} + r^2 d\varphi^2\,,\qquad \phi=\phi(r)\,,
\ee
we find that the functions must satisfy
\be
\alpha (f \phi' (r\phi'-1) - \lambda r e^{-2\phi})(\phi''+ \phi'^2)=0\,.
\ee
Thus there are two branches of solutions.  The first branch, corresponding to the vanishing of the first bracket, is
\be
\phi=\ft12 \log \Big(\fft{\lambda\ell^2}{M} f\Big)\,,\qquad f=r^2\ell^{-2} - M\,.
\ee
Note that this black hole is not asymptotic to the AdS vacuum where $\phi$ should be a constant.  We find that this solution can be promoted to the general BTZ black hole \cite{Banados:1992wn} with both mass $M$ and angular momentum $J$, namely
\begin{equation}
ds^{2}=-\left(  \frac{r^{2}}{\ell^{2}}-M+\frac{J^{2}}{4r^{2}}\right)
dt^{2}+\left(  \frac{r^{2}}{\ell^{2}}-M+\frac{J^{2}}{4r^{2}}\right)
^{-1}dr^{2}+r^{2}\left(  d\phi-\frac{J}{2r^{2}}dt\right)^{2}, \label{btz}
\end{equation}%
and the scalar field becomes
\be
\phi=\ft12\log\Big(\frac{\lambda  r^2 \ell }{\sqrt{M^2 \ell ^2-J^2}}-\ft{1}{2} \lambda  \ell ^2 \Big(\frac{M \ell }{\sqrt{M^2 \ell ^2-J^2}}+1\Big)\Big)\,.\label{btzphi}
\ee
In the extremal limit $J=M\ell$, $\phi$ becomes
\be
\phi=c+\ft12\log(r^2 -\ft12 M\ell^2)\,.
\ee
Note that in this extremal limit, constant shift of the $\phi$ becomes the symmetry of the solution and hence
the constant divergence term in (\ref{btzphi}) in this limit can be ignored.  The scalar field $\phi=\log r$ for $\lambda=0=J$ was obtained in \cite{hkmp}.

The second branch, $\phi''+ \phi'^2$, leads to
\be
\phi=\log \fft{\ell\sqrt{\lambda}}{r_0} (r-r_0)\,,\qquad f= \ell^{-2} r^2\,,\qquad \Lambda_0=-\ell^{-2} - \alpha \ell^{-4}\,.
\ee
When $\Lambda_0=1/(4\alpha)$, this branch contain a new solution
\be
\phi=\log\fft{r-r_0}{a}\,,\qquad f=-\frac{r^2}{2 \alpha } + \frac{r \left(2 \alpha  a^2 \lambda +r_0^2\right)}{\alpha  r_0}-\frac{2 \alpha  a^2 \lambda +r_0^2}{2 \alpha }\,.
\ee
We also find a nearly de Sitter cosmological solution analogous to that in $p=4$, namely
\be
a=e^{H t}\,,\qquad \phi = \log \left(e^{H t} + \sqrt{-\ft{\lambda}{H^2}}\right)\,,
\qquad\Lambda_0=H^2 \left(1-\alpha  H^2\right)\,.
\ee

\subsection{$p=2$}

As in the previous case, we find that for the static metric
\be
ds^2 = -f(r) dt^2 + \fft{dr^2}{f(r)}\,,\label{p2static}
\ee
there is a factorization of the equation
\be
\alpha (-\lambda e^{-2\phi} + f \phi') (\phi'^2 + \phi'')=0\,.
\ee
The first branch requires $\Lambda_0=0$ and we have
\be
f=\fft{\lambda e^{-2\phi}}{\phi'}\,.
\ee
In other words, the equations can be solved by a generic $\phi$. The second branch solution is
\be
\phi = \log\fft{r-r_0}{r_0}\,,\qquad
f=a^2\lambda \pm (r-r_0)^2 \sqrt{\fft{\Lambda_0}{\alpha} (1 - \fft{c}{r-r_0})}\,.
\ee
We find that for the cosmological de Sitter metric, with $\Lambda_0=-\alpha H^4/3$, the scalar $\phi$ has to be a constant, given by $e^{2\phi}=-3\lambda/H^2$. This is exact the de Sitter vacuum.

\section{The $D\rightarrow p+1$ theory}

It was shown \cite{lp} that there exists also $D\rightarrow p+1$ and $D\rightarrow p+2$ limits of EGB gravity using the Kaluza-Klein approach. In this section, we study the $D\rightarrow p+1$ limit of the EGB theory, with $p\le 3$.  The Lagrangian for the $D\rightarrow p+1$ theory is \cite{lp}
\be
{\cal L} =\sqrt{-g}e^{\phi}\Big[R
-2\Lambda_0
+2\alpha\Big(\lambda Re^{-2\phi}-2G^{\mu\nu}\partial_\mu\phi\partial_\nu\phi+(\partial\phi)^2\Box\phi
+2\lambda(\partial\phi)^2e^{-2\phi}+\lambda^2e^{-4\phi}\Big)\Big]\,.\label{theory2}
\ee
The corresponding Gibbons-Hawking surface term is
\bea
S_{\rm surf} &=& 2\int d^{p-1} x \sqrt{-h} \Big(e^{\phi}(1 + 2\alpha \lambda ^{-2\phi})K\nn\\
&&\qquad - 2\alpha e^{\phi} \big(K^{\mu\nu} \partial_\mu \phi\partial_\nu \phi + K n^\mu n^\nu
\partial_\mu \phi \partial_\nu \phi - K (\partial\phi)^2\big)\Big).
\eea
The covariant equations of motion are given in appendix \ref{app:dp1theory}. The maximally symmetric vacua are determined by
\bea
&& (p-1)\lambda_p - 2\alpha \lambda e^{-2\phi_0} \Big( (p-1)^2 \lambda_p + 2 \lambda e^{-2\phi_0}\Big)=0\,,\nn\\
&&\Lambda_0 = \ft12 (p-1)(p-2) \lambda_p (1 + 2\alpha \lambda e^{-2\phi_0} + \alpha \lambda^2 e^{-4\phi_0}\Big)\,.
\eea
Note that if $\lambda=0$, the vacuum cosmological constant $\lambda_p$ is determined solely by the bare cosmological constant $\Lambda_0$.  Furthermore, unlike the previous example, the maximally symmetric vacua exists in all $p$ dimensions, and hence the theory can be promoted to all $p$ dimensions.

First we consider $p=3$ and we have a bifurcation when $\chi=0$ for the static ansatz, namely
\be
\Big(r +\alpha ( 2 f \phi' (r\phi'-2) -2 r \lambda e^{-2\phi})\Big) (\phi'' + \phi'^2)=0\,.
\ee
The first branch requires that $\Lambda_0=-3/(4\alpha)$, in which case, we have
\be
f=\fft{r^2}{2\alpha}-M\,,\qquad \phi=\log \big(c\sqrt{f} + \sqrt{2\alpha\lambda - c^2 M}\big)\,,
\ee
where $c$ is an integration constant. This solution can be promoted to the BTZ metric (\ref{btz}), but with the scalar field given by
\be
\phi=\log \Big(c\sqrt{r^2 -\ft12 M \pm \ft12 \sqrt{M^2 - J^2\ell^{-2}}} \pm
\sqrt{\lambda \pm c^2\sqrt{{M^2 - J^2\ell^{-2}}}}\Big)\,,
\ee
where $\ell^2=2\alpha$. The ``$\pm$'' in front of $\sqrt{{M^2 - J^2\ell^{-2}}}$ should be the same, but independent of the middle ``$\pm$'' above. The plus sign is preferred so that the scalar is regular on the BTZ horizon.

For the second branch, when $\Lambda_0=-3/(4\alpha)$ and $\lambda_3=-1/(2\alpha)$, we have
two solutions:
\be
f=\fft{r^2-r_0^2 + 2a^2 \alpha \lambda}{2\alpha}\,,\qquad \hbox{or}\qquad
f=\fft{(r-r_0)^2 + 2a^2 \alpha \lambda}{2\alpha}\,.
\ee
For both metrics, $\phi=\log(r-r_0)/a$.

For $p=2$, the static ansatz yields
\be
\Big(1+ 2\alpha f \phi'^2 -2\alpha\lambda e^{-2\phi}\Big) (\phi'' + \phi'^2)=0\,.
\ee
Provided with $\Lambda_0=1/(4\alpha)$, all the equations are simply solved by the vanishing of the first bracket.  For the second branch, namely $\phi = \log\big((r-r_0)/r_0\big)$, we have
\be
f=\fft{1}{2\alpha} \Big(2a^2\alpha\lambda+(r-r_0)\sqrt{(1-4\Lambda_0 \alpha)(r-r_0)^2 + c}
-(r-r_0)^2\Big)\,.
\ee

\section{The theory of $D\rightarrow p+2$}

This limit is available only for $p\le 2$ dimensions, and the Lagrangian is \cite{lp}
\bea\label{theory3}
{\cal L}_p&=&\sqrt{-g}e^{2\phi}\Big[R
-2\Lambda_0+2(\partial\phi)^2+2\lambda e^{-2\phi}+2\alpha\Big(2\lambda \phi Re^{-2\phi}-2(\partial\phi)^2\Box\phi\nn\\
&&-((\partial\phi)^2)^2-2\lambda(\partial\phi)^2e^{-2\phi}-\lambda^2e^{-4\phi}\Big)\Big]\,.
\eea
The corresponding Gibbons-Hawking surface term is relatively simply, given by
\be
S_{\rm surf} = 2\int d^{p-1} x \sqrt{-h} (e^{2\phi} + 4\alpha \lambda \phi) K .
\ee
The covariant equations of motion are given in appendix \ref{app:dp2theory}. The maximally symmetric vacua are given by
\bea
&&(p-1)\lambda_p + 2\alpha \lambda^2 e^{-4\phi_0} - \lambda e^{-2\phi_0}
\big(1 + (p-1) \lambda_p (2(p-2)\phi_0-p)\big) =0\,,\nn\\
&&\Lambda_0 = \lambda e^{-2\phi_0} (1- \alpha \lambda e^{-2\phi_0}) +
\ft12 (p-1)(p-2) \lambda_p (1 + 4 \alpha \phi_0 \lambda e^{-2\phi_0})\,.
\eea
When $p=2$, the $\chi=0$ static solution have two branches, characterized by the factorization of the equation
\be
(2\alpha f \phi'^2 -2\alpha\lambda e^{-2\phi} -1) (\phi'' + \phi'^2)=0\,.
\ee
The vanishing of the first bracket can solve all the equations of motion provided that $\Lambda_0=-3/(4\alpha)$.  The metric function $f$ for the the second branch is
\be
f=\fft{1}{2\alpha} (r-r_0)^2 - c \sqrt{r-r_0} +\lambda a^2 \,,
\ee
and $\phi$ is the same as the previous examples.

\section{Holographic properties}

The EGB theory serves as an important gravity model in study the AdS/CFT correspondence. Many of the important holographic properties have been obtained.  The holographic $a$-charge \cite{Myers:2010xs,Myers:2010tj} and the overall coefficients of two-point functions of the energy-momentum tensor $C_T$ \cite{Buchel:2009sk} are given by
\be
a=\ell^{D-2} - 2\alpha (D-2)(D-3) \ell^{D-4}\,,\qquad
C_T=\ell^{D-2} - 2\alpha (D-3)(D-4)\ell^{D-4}\,,
\ee
Note that we have chosen the convention of \cite{Li:2017txk} that strips of inessential overall numerical coefficients such that $a=C_T=\ell^{D-2}$ for Einstein gravity.  They satisfy the differential relation \cite{Li:2017txk}
\be
C_T = \fft{1}{D-2} \ell \fft{\partial a}{\partial \ell}\,.
\ee
(See also \cite{Li:2018drw,Lu:2019urr}.) If we take the limit $D\rightarrow 4$ while keeping $\alpha/(D-4)\rightarrow \alpha$, we have
\be
a=\fft{c}{D-4} + \ell^2 - 4\alpha \log \fft{\ell}{L}\,,\qquad C_T=-2\alpha + \ell^2\,.
\ee
It is clear that the differential relation still holds in this limit.  We would like to verify the above result in the $D=4$ theory.

We first consider the $p=4$ theory (\ref{theory1}), which is expected to be dual to certain strongly coupled CFT in $d=3$ dimensions, where is no conformal anomaly \cite{EntangleE}.  Nevertheless, one can define a sensible $a$-charge that is universal contribution to the entanglement entropy \cite{EntangleE}.  The simplest way to calculate the $a$ charge is to consider the hyperbolic AdS vacuum black hole
\be
ds_p^2= - f dt^2 + \fft{dr^2}{f} + r^2 d\Omega_{p-2,k=-1}^2\,,\qquad f=\fft{r^2}{\ell^2} - 1\,.
\ee
The metric is maximally supersymmetric and yet it contains a horizon at $r_+=\ell$.  For $p=4$, we have a running scalar with
\be
\phi=\log \Big(\fft{\ell}{L} + \fft{\sqrt{(L^2\lambda -1)(r^2-\ell^2)}}{L}\Big)\,,
\ee
We thus have $\phi(r_+)=\log(\ell/L)$. The Iyer-Wald entropy can be easily obtained and we have
\be
a=\ell^2 - 4\alpha \log\fft{\ell}{L}\,.
\ee
Note that the logarithmic dependence arises from the $\phi E^{\rm GB}$ term.  The overall coefficient of the holographic two-point function of the energy-momentum tensor is proportional to $\ell^2 \kappa_{\rm eff}$ given by (\ref{p4kappaeff}) with $\lambda_4=-1/\ell^2$.  Thus we recover the $D\rightarrow 4$ limit of the $a$ and $C_T$, with the differential relation remains valid.  It is important that we must consider the scalar hairy black hole in order to obtain the right logarithmic contribution to the $a$-charge, another indication that the scalar is indispensable in the four-dimensional theory.

There are two $p=3$ theories, given by (\ref{theory1}) and (\ref{theory2}).  At the linearized level, the effective coupling that is the inverse of the Newton's constant are $\kappa_{\rm eff}= 1-2\lambda e^{-2\phi_0}$ and $\kappa_{\rm eff}= e^{\phi_0} + \lambda e^{-\phi_0}$ respectively. In each case, we have $a= C_T\sim \ell \kappa_{\rm eff}$.

\section{Conclusion}

In this paper, we study the lower dimensional limits of EGB gravities.  Three action principles were explicitly constructed in \cite{lp}. The minimum theories all consist of the metric as well as a scalar field that is of the Horndeski type. We examined the vacua of these theories and also obtained some exact solutions that describes static black holes and also cosmology.

We found that there are two types of vacua.  One is maximally symmetric and the scalar field is a constant.  The effective cosmological constants of the vacua arise as the root of a quadratic equation, and intriguingly the reality condition of the scalar will select only one of them, when the cosmological constants have the opposite signs.  We found that in the $p=4$ theory, the kinetic term of the linearized scalar equation vanishes identically, so that the linear spectrum involves only a massless graviton. This has important consequence in subject of gravitational waves.  It can be established that there are no news associated with the scalar in the Bond-Sachs framework, but the effects of the Gauss-Bonnet term are quite significant in the sense that they arise just one order after the integration constants and also arise in the quadrupole of the gravitational source \cite{lm}.

There are also nearly maximally symmetric vacua where the scalar takes nontrivial coordinate dependence and hence breaks the full maximal symmetry.  This implies that for matter coupled to the metric only will respect the Lorentz symmetry whilst matter coupled to the scalar will experience the Lorentz violation.  We also obtained black holes and cosmological solutions with nontrivial scalar hair.  In particular, we obtained exact scalar hairy BTZ black holes in three dimensions.  For asymptotic AdS spaces, we also studied the holographic properties associated with the $a$-charges and the overall coefficient of the two-point functions of the energy-momentum tensor. The richness and relative simplicity of these theories make them interesting to investigate further.

\section*{Acknowledgement}

We are grateful to Yue-Zhou Li, Chun-Shan Lin, Hai-Shan Liu, Yi Pang, Zhao-Long Wang, Jun-Bao Wu, Run-Qiu Yang for useful discussions. The work is supported in part by NSFC (National Natural Science Foundation of China) Grants No. 11875200 and No. 11935009.

\appendix
\section{Covariant equations of motion}

\subsection{The $D\rightarrow p$ theory}
\label{app:dptheory}

The Lagrangian of this theory is given in (\ref{theory1}) and the action can be written as
\bea
S &=& \int d^p x \sqrt{-g} (L_0 + L_1 + L_2 + L_3 + L_4)\,,\qquad L_0 =  R - 2\Lambda_0\,,\nn\\
L_1 &=& \alpha \phi E^{\rm GB}\,,\qquad \qquad L_2 = 4\alpha G^{\mu\nu} \partial_\mu\phi\partial_\nu \phi\,,\qquad L_3 = -2\alpha \lambda e^{-2\phi} R\,,\nn\\
L_4 &=& -4\alpha\left(  \partial\phi\right)  ^{2}\square\phi+2\alpha((\partial\phi)^{2})^{2} -12\alpha\lambda\left(\partial\phi\right)  ^{2} e^{-2\phi}-6\alpha\lambda^{2}e^{-4\phi}
\eea
The Einstein equation of motion is given by
\be
E_{ab} \equiv E_{0ab} + E_{1ab}+ E_{2ab}+ E_{3ab}+ E_{4ab} =0\,.\label{einsteincoveom1}
\ee
Here $E_{i ab}$ corresponds to the metric variation of $\sqrt{-g} L_i$ and they are
\bea
E_{0ab} &=& G_{ab}+\Lambda_{0}g_{ab}\,,\nn\\
E_{1ab} &=&\alpha\phi H_{ab}+2\alpha\Big[4R_{c(a}\nabla^{c}\nabla_{b)}\phi  -R\nabla_{a}\nabla_{b}%
\phi-2G_{ab}\square\phi +2\left(  R_{acbd}-g_{ab}R_{cd}\right)  \nabla^{c}\nabla^{d}\phi\Big]\,,\nn\\
H_{ab} &=& 2\left(R_{ab}R  -2R_{a}{}^{c}R_{bc}-2R^{cd}R_{acbd}+R_{a}{}^{cde}R_{bcde}\right)  -\ft{1}{2}g_{ab}E^{\rm GB}\,,\nn\\
E_{2ab} &=& 2\alpha\Big[4R_{(a}{}^{c}\partial_{b)}\phi\partial_{c}\phi -2\left(  \nabla_{a}\nabla_{b}\phi\right) \square\phi-R\,\partial_{a}\phi\partial_{b}\phi +2\left(  \nabla_{c}\nabla_{a}\phi\right)  \left(  \nabla^{c}
\nabla_{b}\phi\right)\nn\\
&&-G_{ab}\left(\partial\phi\right)^{2} +g_{ab}\big(\left(  \square\phi\right)^{2} -2R^{cd}\partial_{c}\phi\partial_{d}\phi-\left(\nabla_{c}\nabla_{d}%
\phi\right)(\nabla^{c}\nabla^{d}\phi)\big)+2R_{a}{}^c{}_b{}^d%
\partial_{c}\phi\partial_{d}\phi\Big]\,,\nn\\
E_{3ab} &=&-\alpha\lambda e^{-2\phi}\left[  2R_{ab}-8\partial_{a}\phi\partial
_{b}\phi+4\nabla_{a}\nabla_{b}\phi-g_{ab}\left(  R+4\square\phi-8\left(
\partial\phi\right)  ^{2}\right)  \right],\nn\\
E_{4ab} &= & 4\alpha\Big[2\left(\nabla^{c}\nabla_{(a}\phi\right)  \partial_{b)}\phi
\,\partial_{c}\phi -\partial_{a}\phi\partial_{b}\phi\,\square\phi -g_{ab}(\nabla^{c}\nabla^{d}\phi) \partial_{c}\phi\partial_{d}\phi\Big]
+3\alpha\lambda^{2} e^{-4\phi}g_{ab}\nn\\
&&
+\alpha\left[  4\partial_{a}\phi\partial_{b}\phi\,(
\partial\phi)^{2}-g_{ab}((  \partial\phi)
^{2})  ^{2}\right]  -12\alpha\lambda e^{-2\phi}\left[
\partial_{a}\phi\partial_{b}\phi  -\ft12 g_{ab}\left(  \partial\phi\right)
^{2}\right],\label{einsteincoveom2}
\eea
where it is understood that $H_{ab}$ vanishes for $p\le 4$.  The scalar equation is
\bea
&&  \alpha\Big\{  4\left[  2\left(  \square\phi\right)  ^{2}+R\square
\phi\right]  +4\lambda e^{-2\phi}\left[  R+6\square\phi-6\left(  \partial
\phi\right)  ^{2}\right] -8\left[R^{ab}+2(\nabla^{a}\nabla^{b}\phi)  \right]
\partial_{a}\phi\partial_{b}\phi\nn\\
&&-8\left[  R_{ab}+\left(  \nabla_{a}\nabla_{b}\phi\right)  \right]\nabla^{a}\nabla^{b}\phi
+24\lambda^{2}e^{-4\phi}-8\left(  \partial\phi\right)  ^{2}%
\square\phi + E^{\rm GB}\Big\} =0\,.\label{scalarcoveom}
\eea
The trace of the Einstein equation (\ref{einsteincoveom1}) is
\bea
&&  +2\left[  R+p\left(  \Lambda_{0}-\ft{1}{2}R\right)  \right] +
2\alpha\Big\{  2\left[  \left(  p-2\right)  \left(  \square\phi\right)^{2}
+\left(  p-3\right)  R\square\phi\right]\nn\\
&&  +\lambda e^{-2\phi}\left[\left(  p-2\right)  R+4\left(  p-1\right)  \square\phi-2\left(  p+2\right)
\left(  \partial\phi\right)  ^{2}\right]\nn\\
&&-4\left[  \left(  p-3\right)  R_{ab}+\left(  p-2\right)  \left(
\nabla_{b}\nabla_{a}\phi\right)  \right]  \partial^{a}\phi\partial^{b}%
\phi \nn\\
&&-2\left[  2\left(  p-3\right)  R_{ab}+\left(  p-2\right)  \left(
\nabla_{b}\nabla_{a}\phi\right)  \right]\nabla^{a}\nabla^{b}\phi  +3p\lambda^{2}e^{-4\phi}-4\left(  \partial\phi\right)  ^{2}%
\square\phi\Big\} \nonumber\\
&&  -\alpha\left(  p-4\right)  \left[  \phi E^{\rm GB}-2\left(
\partial\phi\right)  ^{2}R+2((  \partial\phi)  ^{2})^{2}\right]=0\,, \label{tracecoveom}
\eea
Note that the last term drops out for $p=4$, in which case we can derive the purely gravitational scalar constrain equation (\ref{p4cons}) from (\ref{scalarcoveom}) and (\ref{tracecoveom}).

\subsection{The $D\rightarrow p+1$ theory}
\label{app:dp1theory}

The Lagrangian of this theory is given in (\ref{theory2}) and the action can be written as
\bea
S &=& \int d^p x \sqrt{-g} (L_0 + L_1 + L_2 + L_3 + L_4)\,,\qquad L_0 =  e^{\phi}(R - 2\Lambda_0)\,,\nn\\
L_1 &=& -4\alpha e^{\phi} G^{\mu\nu} \partial_\mu\phi\partial_\nu \phi\,,\qquad L_2 = 2\alpha \lambda e^{-\phi} R\,,\nn\\
L_3 &=& 2\alpha e^{\phi} \Big(\left(  \partial\phi\right)^{2}\square\phi+ 2\lambda\left(\partial\phi\right)  ^{2} e^{-2\phi}+\lambda^{2}e^{-4\phi}\Big),
\eea
The Einstein equation of motion is given by
\be
E_{ab} \equiv E_{0ab} + E_{1ab}+ E_{2ab}+ E_{3ab} =0\,.\label{dp1einsteincoveom1}
\ee
Here $E_{i ab}$ corresponds to the metric variation of $\sqrt{-g} L_i$ and they are
\bea
E_{0ab}&=&e^{\phi}\left\{  R_{ab}-\ft{1}{2}g_{ab}R+\Lambda_{0}g_{ab}%
-\nabla_{a}\nabla_{b}\phi-\partial_{a}\phi\partial_{b}\phi +
g_{ab}\big[  \square\phi+\left(  \partial
\phi\right)  ^{2}\big]\right\},\nn\\
E_{1ab}  &=& -\alpha e^{\phi}\Big\{  -2\nabla_{a}\nabla_{b}\phi\, \big[
2\square\phi+\left(  \partial\phi\right)  ^{2}\big]  -2R_{ab}\left(
\partial\phi\right)  ^{2} -2\left(  \square\phi+R\right)  \partial_{a}\phi\partial_{b}\phi \nonumber\\
&&  +g_{ab}\left[  -2\left(  \nabla_{c}\nabla_{d}\phi\right)  \big(
\nabla^{c}\nabla^{d}\phi\big)  +2\square\phi\big[  \square\phi+\left(
\partial\phi\right)  ^{2}\big]  +\left(  \partial\phi\right)  ^{2}R\right]
 \nonumber\\
&&
+4  \nabla_{c}\nabla_{a}\phi\, \nabla^{c}\nabla_{b}\phi  +4\nabla^{c}\nabla_{(a}\phi\,\partial_{b)}\phi\,\partial_{c}\phi
+8R^{c}{}_{(a}\partial_{b)}\phi\,\partial_{c}\phi \nonumber\\
&& +8R_{a}{}^{c}{}_b{}^{d}\,\partial_{c}\phi\partial_{d}\phi-4g_{ab}\big[  2R_{cd}+\big(
\nabla^{c}\nabla^{d}\phi\big)\big]  \partial_{c}\phi\partial_{d}%
\phi \Big\},\nn\\
E_{2ab}&=&2\alpha\lambda e^{-\phi} \left\{  R_{ab}-\ft{1}{2}g_{ab}R+\nabla
_{a}\nabla_{b}\phi-\partial_{a}\phi\partial_{b}\phi-g_{ab}\big[\square\phi-\left(  \partial\phi\right)
^{2}\big]\right\},\nn\\
E_{3ab}  &=& -\alpha e^{\phi}\Big\{  4\left(  \nabla^{c}\nabla
_{(a}\phi\right)  \partial_{b)}\phi\,\partial_{c}\phi  -2  \big[\square\phi-\left(  \partial\phi\right)
^{2} + 2 \lambda e^{-2\phi}\big]  \partial_{a}\phi\partial_{b}\phi\nn\\
&&+g_{ab}\left[  \lambda
^{2} e^{-4\phi}+2\lambda e^{-2\phi}\left(  \partial\phi\right)  ^{2}-\big(
(  \partial\phi)  ^{2}\big)  ^{2}-2\big(  \nabla^{c}\nabla^{d}\phi\big)  \partial_{c}\phi\partial_{d}\phi\right] \Big\}\,.
\eea
The scalar equation is given by
\bea
\!\!\!&&R-2\Lambda_{0} + 2\alpha \Big\{4R^{ab}\big(  \nabla_{a}\nabla_{b}\phi
+\partial_{a}\phi\partial_{b}\phi\big) -\lambda e^{-2\phi}\big[  R-2\left(
\partial\phi\right)  ^{2}+4\square\phi\big]-2\left(  \square\phi\right)^{2}\nn\\
\!\!\!&&-R\big[  2\square\phi+\left(  \partial\phi\right)  ^{2}\big]
+\big((  \partial\phi)  ^{2}\big)  ^{2}+2\left[  \nabla_{a}\nabla_{b}\phi+2\partial
_{a}\phi\partial_{b}\phi\right]  \big(  \nabla^{a}\nabla^{b}\phi\big)-3\lambda^{2}e^{-4\phi}
\Big\}=0.
\eea
In $p=3$ dimensions, the trace of the Einstein equation and scalar equation yield
\begin{equation}
4\Lambda_{0}-R+2\left(  1+\alpha R\right)  \big[
\square\phi+\left(  \partial\phi\right)  ^{2}\big]  -4\alpha R^{ab}\big(
\nabla_{a}\nabla_{b}\phi+\partial_{a}\phi\partial_{b}\phi\big) =0\,.
\end{equation}

\subsection{The $D\rightarrow p+2$ theory}
\label{app:dp2theory}

The Lagrangian of this theory is given in (\ref{theory3}) and the action can be written as
\bea
S &=& \int d^p x \sqrt{-g} (L_0 + L_1 + L_2)\,,\nn\\
L_0 &=&  e^{2\phi}(R - 2\Lambda_0)\,,\qquad L_1 = 4\alpha \lambda R\,,\nn\\
L_2 &=&2e^{2\phi}\big[  \left(  \partial\phi\right)  ^{2}+\lambda
e^{-2\phi}\big]\nn\\
&&  -2\alpha e^{2\phi}\left[2\left(  \partial\phi\right)  ^{2}%
\square\phi+\big((  \partial\phi)  ^{2}\big)  ^{2} +2\lambda\left(  \partial\phi\right)  ^{2}e^{-2\phi}+\lambda^{2}e^{-4\phi}\right].
\eea
The Einstein equation of motion is given by
\be
E_{ab} \equiv E_{0ab} + E_{1ab}+ E_{2ab}+ E_{3ab} =0\,.\label{dp2einsteincoveom1}
\ee
Here $E_{i ab}$ corresponds to the metric variation of $\sqrt{-g} L_i$ and they are
\bea
E_{0ab} &=& e^{2\phi}\left\{  R_{ab}-\ft{1}{2}g_{ab}R+\Lambda_{0}g_{ab}%
-2\nabla_{a}\nabla_{b}\phi -4\partial_{a}\phi\partial_{b}\phi
+2g_{ab}\big(  \square\phi+2\left(  \partial\phi\right)  ^{2}\big) \right\},\nn\\
E_{1ab}&=&4\alpha\lambda\left\{  \phi\left(  R_{ab}-\ft{1}{2}g_{ab}R\right)
-\nabla_{a}\nabla_{b}\phi+g_{ab}\square\phi\right\}\,,\nn\\
E_{2ab}  &= &e^{2\phi}\Big\{  8\alpha\left(  \nabla
^{c}\nabla_{(a}\phi\right)  \partial_{b)}\phi\,\partial_{c}\phi +2\big[1-2\alpha ( \square\phi-\left(\partial\phi\right)^{2}+\lambda e^{-2\phi})\big] \partial_{a}\phi\partial_{b}\phi \nonumber\\
&&  -g_{ab}\Big[\left(\partial \phi\right)^{2}\big[  1+3\alpha\left(  \partial\phi\right)  ^{2}\big]+\alpha\big(\nabla^{c}\nabla^{d} \phi\big)  \partial_{c}\phi\partial_{d}\phi\nn\\
&&\qquad\qquad+\lambda e^{-2\phi}\big[  1-2\alpha\left(  \partial \phi\right)  ^{2}\big] -\alpha\lambda^{2}e^{-4\phi}\Big]\Big\}.
\eea
The scalar equations of motion is
\bea
&&R -  2\big(\Lambda_{0}+\square\phi+\left(  \partial\phi\right)  ^{2}\big)+
2\alpha \Big\{2\left(\square
\phi\right)^{2}-\big((\partial\phi)^{2}\big)^{2}+2\left(  \partial\phi\right)  ^{2}\square\phi
-2R^{ab}\partial_{a}\phi\partial_{b}\phi\nn\\
&& - 2(  \nabla^{a}\nabla^{b}\phi)  \left[
\nabla_{a}\nabla_{b}\phi+2\partial_{a}\phi\partial_{b}\phi\right]+\lambda e^{-2\phi}\left(  R+2\square\phi\right)+ \lambda^{2} e^{-4\phi}\Big\}=0\,.
\eea


\begin{thebibliography}{99}

\bibitem{Stelle:1976gc}
K.S.~Stelle,
``Renormalization of higher derivative quantum gravity,''
Phys. Rev. D \textbf{16}, 953-969 (1977)
doi:10.1103/PhysRevD.16.953
``Classical gravity with higher derivatives,''
Gen. Rel. Grav. \textbf{9}, 353-371 (1978)
doi:10.1007/BF00760427

\bibitem{L1}
  D.~Lovelock,
  ``The Einstein tensor and its generalizations,''
  J.\ Math.\ Phys.\  {\bf 12} (1971) 498.

\bibitem{H}
 G.W.~Horndeski,
``Second-order scalar-tensor field equations in a four-dimensional space,''
 Int.\ J.\ Theor.\ Phys.\  {\bf 10} (1974) 363.

\bibitem{Deffayet:2009mn}
  C.~Deffayet, S.~Deser and G.~Esposito-Farese,
  ``Generalized Galileons: All scalar models whose curved background extensions maintain second-order field equations and stress-tensors,''
  Phys.\ Rev.\ D {\bf 80} (2009) 064015,
  arXiv:0906.1967 [gr-qc];
C.~Deffayet, X.~Gao, D.A.~Steer and G.~Zahariade,
  ``From k-essence to generalised Galileons,'' {\em Phys.\ Rev.\ D }{\bf 84}, (2011) 064039,
  doi:10.1103/PhysRevD.84.064039, arXiv:1103.3260 [hep-th].

\bibitem{VanAcoleyen:2011mj}
  K.~Van Acoleyen and J.~Van Doorsselaere,
  ``Galileons from Lovelock actions,''
  Phys.\ Rev.\ D {\bf 83} (2011) 084025,
  arXiv:1102.0487 [gr-qc];
  C.~Charmousis, B.~Gouteraux and E.~Kiritsis,
  ``Higher-derivative scalar-vector-tensor theories: black holes, Galileons, singularity cloaking and holography,''
  JHEP {\bf 1209} (2012) 011,varXiv:1206.1499 [hep-th];
  C.~Charmousis,
  ``From Lovelock to Horndeski's generalized scalar tensor theory,''
  Lect.\ Notes Phys.\  {\bf 892} (2015) 25, arXiv:1405.1612 [gr-qc].

\bibitem{GL}
D.~Glavan, C.~Lin
``Einstein-Gauss-Bonnet gravity in 4-dimensional space-time,"
Phys.\ Rev.\ Lett. {\bf 124}, 081301 (2020).

\bibitem{lp} H.~L\"u and Y.~Pang,
``Horndeski gravity as $D\rightarrow4$ limit of Gauss-Bonnet,''
  arXiv:2003. 11552 [gr-qc].

\bibitem{Kobayashi:2020wqy}
  T.~Kobayashi,
``Effective scalar-tensor description of regularized Lovelock gravity in four dimensions,''
  arXiv:2003.12771 [gr-qc].

\bibitem{Gurses:2020ofy}
  M.~Gurses, T.C.~Sisman and B.~Tekin,
  ``Is there a novel Einstein-Gauss-Bonnet theory in four dimensions?,''
  arXiv:2004.03390 [gr-qc].

\bibitem{Ai:2020peo}
  W.Y.~Ai, ``A note on the novel 4D Einstein-Gauss-Bonnet gravity,''
  arXiv:2004.02858 [gr-qc].

\bibitem{Shu:2020cjw}
  F.W.~Shu,
  ``Vacua in novel 4D Einstein-Gauss-Bonnet gravity: pathology and instability?''
  arXiv:2004.09339 [gr-qc].

\bibitem{backgrounds}
R.~Konoplya and A.~Zinhailo,
arXiv:2003.01188 [gr-qc];
 M.~Guo and P.C.~Li,
  arXiv:2003.02523 [gr-qc];
P.G.S.~Fernandes,
  arXiv:2003.05491 [gr-qc];
R.A.~Konoplya and A.~Zhidenko,
  arXiv:2003.07788 [gr-qc];
S.W.~Wei and Y.X.~Liu,
  arXiv:2003.07769 [gr-qc];
A.~Casalino, A.~Colleaux, M.~Rinaldi and S.~Vicentini,
  arXiv:2003.07068 [gr-qc];
R.~Kumar and S.G.~Ghosh,
  arXiv:2003.08927 [gr-qc];
K.~Hegde, A.N.~Kumara, C.L.A.~Rizwan, A.K.M. and M.S.~Ali,
  arXiv:2003.08778 [gr-qc];
D.D.~Doneva and S.S.~Yazadjiev,
  arXiv:2003.10284 [gr-qc];
S.G.~Ghosh and S.D.~Maharaj,
  arXiv:2003.09841 [gr-qc];
C.-Y. Zhang, P.-C. Li, and M.~Guo,
arXiv:2003.13068 [hep-th];
S.-W. Wei and Y.-X. Liu,
arXiv:2003.14275 [gr-qc];
M.~Churilova,
 arXiv:2004.00513 [gr-qc];
S.U.~Islam, R.~Kumar, and S.G.~Ghosh,
arXiv:2004.01038 [gr-qc];
A.K.~Mishra,
arXiv:2004.01243 [gr-qc];
S.-L.~Li, P.~Wu, and H.~Yu,
arXiv:2004.02080 [gr-qc];
M.~Heydari-Fard, M.~Heydari-Fard, and H.~Sepangi,
arXiv:2004.02140 [gr-qc];
R.A. Konoplya and A.F. Zinhailo,
arXiv:2004.02248 [gr-qc];
 C.Y.~Zhang, S.J.~Zhang, P.C.~Li and M.~Guo,
  arXiv:2004.03141 [gr-qc];
A.~Naveena Kumara, C.L.A.~Rizwan, K.~Hegde, M.S.~Ali and A.K.~M,
  ``Rotating 4D Gauss-Bonnet black hole as particle accelerator,''
  arXiv:2004.04521 [gr-qc];
S.-J.~Yang, J.-J.~Wan, J.~Chen, J.~Yang, and Y.-Q.~Wang,
arXiv:2004.07934 [gr-qc];
A.~Casalino and L.~Sebastiani,
arXiv:2004.10229 [gr-qc];
X.~Zeng, H.~Zhang and H.~Zhang,
[arXiv:2004.12074 [gr-qc]];
X.~Ge and S.~Sin,
[arXiv:2004.12191 [hep-th]];
 G.~Alkac and D.~O.~Devecioglu,
 ``Three dimensional modified gravities as Holographic limits of Lancsoz-Lovelock theories,''
  arXiv:2004.12839 [hep-th];
R.~Kumar, S.U.~Islam and S.G.~Ghosh,
[arXiv:2004.12970 [gr-qc]];
J.~Arrechea, A.~Delhom and A.~Jim\'enez-Cano,
[arXiv:2004.12998 [gr-qc]];
S.G.~Ghosh and S.D.~Maharaj,
  ``Noncommutative inspired black holes in regularised 4D Einstein-Gauss-Bonnet theory,''
  arXiv:2004.13519 [gr-qc];
M.S.~Churilova,
``Quasinormal modes of the test fields in the novel 4D Einstein-Gauss-Bonnet-de Sitter gravity,''
  arXiv:2004.14172 [gr-qc].

\bibitem{2d1}
R.B.~Mann and S.F.~Ross,
``The $D\rightarrow 2$ limit of general relativity,'' Class.\ Quant.\ Grav. {\bf 10}, (1993) 1405, arXiv:gr-qc/9208004.

\bibitem{2d2}
S.~Nojiri and S.D.~Odintsov, ``{Novel cosmological and black hole solutions in Einstein and higher-derivative gravity in two dimensions},'' arXiv:2004.01404 [hep-th].

\bibitem{eoms}
P.G. Fernandes, P.~Carrilho, T.~Clifton, and D.J.~Mulryne,
``Derivation of regularized field equations for the Einstein-Gauss-Bonnet theory in four dimensions,''
  arXiv:2004.08362 [gr-qc]; 

\bibitem{Hennigar:2020lsl}
  R.A.~Hennigar, D.~Kubiznak, R.B.~Mann and C.~Pollack,
``On taking the $D\to 4$ limit of Gauss-Bonnet Gravity: theory and solutions,''
  arXiv:2004.09472 [gr-qc].

\bibitem{Bonifacio:2020vbk}
  J.~Bonifacio, K.~Hinterbichler and L.A.~Johnson,
  ``Amplitudes and 4D Gauss-Bonnet theory,''
  arXiv:2004.10716 [hep-th].

\bibitem{Gibbons:1976ue}
  G.W.~Gibbons and S.W.~Hawking,
  ``Action integrals and partition functions in quantum gravity,''
  Phys.\ Rev.\ D {\bf 15}, 2752 (1977).
  doi:10.1103/PhysRevD.15.2752

\bibitem{Liu:2008zf}
  J.T.~Liu and W.A.~Sabra,
  ``Hamilton-Jacobi counterterms for Einstein-Gauss-Bonnet gravity,''
  Class.\ Quant.\ Grav.\  {\bf 27}, 175014 (2010)
  doi:10.1088/0264-9381/27/17/175014
  [arXiv:0807.1256 [hep-th]].
  H.S.~Liu, H.~L\"u and C.~N.~Pope,
  ``Holographic heat current as Noether current,''
  JHEP {\bf 1709}, 146 (2017)
  doi:10.1007/JHEP09(2017)146
  [arXiv:1708.02329 [hep-th]].

\bibitem{Li:2018rgn}
  Y.Z.~Li, H.~L\"u and H.Y.~Zhang,
  ``Scale invariance vs. conformal invariance: holographic two-point functions in Horndeski gravity,''
  Eur.\ Phys.\ J.\ C {\bf 79}, no. 7, 592 (2019)
  doi:10. 1140/epjc/s10052-019-7096-6
  [arXiv:1812.05123 [hep-th]].

\bibitem{Fan:2016zfs}
  Z.Y.~Fan, B.~Chen and H.~L\"u,
  ``Criticality in Einstein-Gauss-Bonnet gravity: gravity without graviton,''
  Eur.\ Phys.\ J.\ C {\bf 76}, no. 10, 542 (2016)
  doi:10.1140/epjc/s10052-016-4389-x
  [arXiv:1606.02728 [hep-th]].

\bibitem{Banados:1992wn}
  M.~Banados, C.~Teitelboim and J.~Zanelli,
  ``The black hole in three-dimensional space-time,''
  Phys.\ Rev.\ Lett.\  {\bf 69}, 1849 (1992)
  doi:10.1103/PhysRevLett.69.1849
  [hep-th/9204099].

\bibitem{hkmp}
R.A.~Hennigar, D.~Kubiznak, R.B.~Mann and C.~Pollack,
[arXiv:2004.12995 [gr-qc]];


\bibitem{Myers:2010xs}
R.C.~Myers and A.~Sinha,
``Seeing a $c$-theorem with holography,''
Phys.\ Rev.\ D {\bf 82} (2010) 046006
doi:10.1103/PhysRevD.82.046006
[arXiv:1006.1263 [hep-th]].

\bibitem{Myers:2010tj}
R.C.~Myers and A.~Sinha,
``Holographic $c$-theorems in arbitrary dimensions,''
JHEP {\bf 1101} (2011) 125
doi:10.1007/JHEP01(2011)125
[arXiv:1011.5819 [hep-th]].

\bibitem{Buchel:2009sk}
  A.~Buchel, J.~Escobedo, R.C.~Myers, M.F.~Paulos, A.~Sinha and M.~Smolkin,
``Holographic GB gravity in arbitrary dimensions,''
  JHEP {\bf 1003}, 111 (2010)
  doi:10.1007/ JHEP03(2010)111
  [arXiv:0911.4257 [hep-th]].

\bibitem{Li:2017txk}
  Y.Z.~Li, H.~L\"u and J.B.~Wu,
  ``Causality and $a$-theorem Constraints on Ricci polynomial and Riemann cubic gravities,''
  Phys.\ Rev.\ D {\bf 97}, no. 2, 024023 (2018)
  doi:10.1103/Phys RevD.97.024023
  [arXiv:1711.03650 [hep-th]].

\bibitem{Li:2018drw}
  Y.Z.~Li, H.~L\"u and Z.F.~Mai,
  ``Universal structure of covariant holographic two-point functions in massless higher-order gravities,''
  JHEP {\bf 1810}, 063 (2018)
  doi:10.1007/ JHEP10(2018)063
  [arXiv:1808.00494 [hep-th]].

\bibitem{Lu:2019urr}
  H.~L\"u and R.~Wen,
``Holographic $(a,c)$-charges and their universal relation in $d=6$ from massless higher-order gravities,''
  Phys.\ Rev.\ D {\bf 99}, no. 12, 126003 (2019)
  doi:10.1103/ PhysRevD.99.126003
  [arXiv:1901.11037 [hep-th]].

\bibitem{EntangleE}
  C.~Imbimbo, A.~Schwimmer, S.~Theisen and S.~Yankielowicz,
  ``Diffeomorphisms and holographic anomalies,''
  Class.\ Quant.\ Grav.\  {\bf 17}, 1129 (2000)
  doi:10.1088/0264-9381/ 17/5/322
  [hep-th/9910267];
  L.Y.~Hung, R.C.~Myers and M.~Smolkin,
  ``On holographic entanglement entropy and higher curvature gravity,''
  JHEP {\bf 1104}, 025 (2011)
  doi:10.1007/ JHEP04(2011)025
  [arXiv:1101.5813 [hep-th]].
  
\bibitem{lm}
H.~L\"u and P.J.~Mao, ``Asymptotic structure of Einstein-Gauss-Bonnet theory in lower dimensions,''
to appear.

\end{thebibliography}
\end{document}